\documentstyle[12pt]{article}
\begin{document}
\def\beq{\begin{equation}}
\def\eeq{\end{equation}}
\def\bea{\begin{eqnarray}}
\def\eea{\end{eqnarray}}
\def\ve{\vert}
\def\vel{\left|}
\def\ver{\right|}
\def\nnb{\nonumber}
\def\ga{\left(}
\def\dr{\right)}
\def\aga{\left\{}
\def\adr{\right\}}
\def\rar{\rightarrow}
\def\nnb{\nonumber}
\def\la{\langle}
\def\ra{\rangle}
\def\ba{\begin{array}}
\def\ea{\end{array}}
\def\tep{$B \rar K \ell^+ \ell^-$}
\def\tepm{$B \rar K \mu^+ \mu^-$}
\def\tept{$B \rar K \tau^+ \tau^-$}
\def\ds{\displaystyle}


\renewcommand{\textfraction}{0.2}    
\renewcommand{\topfraction}{0.8}   
\renewcommand{\bottomfraction}{0.4}   
\renewcommand{\floatpagefraction}{0.8}
\newcommand\mysection{\setcounter{equation}{0}\section}

\def\baeq{\begin{appeq}}     \def\eaeq{\end{appeq}}  
\def\baeeq{\begin{appeeq}}   \def\eaeeq{\end{appeeq}}
\newenvironment{appeq}{\beq}{\eeq}   
\newenvironment{appeeq}{\beeq}{\eeeq}
\def\bAPP#1#2{
 \markright{APPENDIX #1}
 \addcontentsline{toc}{section}{Appendix #1: #2}
 \medskip
 \medskip
 \begin{center}      {\bf\LARGE Appendix  :}{\quad\Large\bf #2}
\end{center}
 \renewcommand{\thesection}{#1.\arabic{section}}
\setcounter{equation}{0}
        \renewcommand{\thehran}{#1.\arabic{hran}}
\renewenvironment{appeq}
  {  \renewcommand{\theequation}{#1.\arabic{equation}}
     \beq
  }{\eeq}
\renewenvironment{appeeq}
  {  \renewcommand{\theequation}{#1.\arabic{equation}}
     \beeq
  }{\eeeq}
\nopagebreak \noindent}

\def\eAPP{\renewcommand{\thehran}{\thesection.\arabic{hran}}}

\renewcommand{\theequation}{\arabic{equation}}
\newcounter{hran}
\renewcommand{\thehran}{\thesection.\arabic{hran}}

\def\bmini{\setcounter{hran}{\value{equation}}
\refstepcounter{hran}\setcounter{equation}{0}
\renewcommand{\theequation}{\thehran\alph{equation}}\begin{eqnarray}}
\def\bminiG#1{\setcounter{hran}{\value{equation}}
\refstepcounter{hran}\setcounter{equation}{-1}
\renewcommand{\theequation}{\thehran\alph{equation}}
\refstepcounter{equation}\label{#1}\begin{eqnarray}}


\newskip\humongous \humongous=0pt plus 1000pt minus 1000pt
\def\caja{\mathsurround=0pt}
\def\eqalign#1{\,\vcenter{\openup1\jot
\caja   \ialign{\strut \hfil$\displaystyle{##}$&$
\displaystyle{{}##}$\hfil\crcr#1\crcr}}\,}



\def\bos{\lower 0.7cm\hbox{{\vrule width 0pt height 1.5cm}}}
\def\aaa{\lower 0.cm\hbox{{\vrule width 0pt height .8cm}}}
\def\dol{\lower 0.6cm\hbox{{\vrule width 0pt height .8cm}}}


\title{ {\small {\bf RARE RADIATIVE 
$B \rar \tau^+ \tau^- \gamma$ DECAY } } }

\author{ {\small T. M. AL\.{I}EV$^1$ \,,
N. K. PAK$^2$\,\, and \,\, 
M. SAVCI$^2$}\\
{\small 1) Physics Department, Girne American University} \\
{\small Girne, Cyprus}\\
{\small 2) Physics Department, Middle East Technical University} \\
{\small 06531 Ankara, Turkey}}

\date{}

\begin{titlepage}
\maketitle
\thispagestyle{empty}

\begin{abstract}
\baselineskip  0.7cm

The radiative $B \rar \tau^+ \tau^- \gamma$ decay is investigated in the
framework of the Standard Model. When only short (short and long together) 
distance contributions are taken into account, the Branching Ratio is found
as $9.54 \times 10^{-9}$ ($1.52  \times 10^{-8}$), for the value of the cut 
$\delta = 0.01$ imposed on the photon energy.

\end{abstract}

\vspace{1cm}
\end{titlepage}

\section{Introduction}
Experimental discovery of the inclusive and exclusive $B \rar X_s \gamma$
and $B \rar K^* \gamma$ \cite{R1} decays stimulated the study of the
radiative rare $B$ meson decays with a new momentum. From theoretical point
of view this is due to the fact that they are very sensitive to the flavor
structure of the electroweak interactions, as well as $QCD$ radiative
corrections and the new physics beyond the Standard Model (SM) \cite{R2}.
From experimental point of view studying radiative $B$--meson decays allows
more precise determination of the parameters of the SM, such as the elements
of the Cabibbo--Kobayashi--Maskawa (CKM) matrix, the leptonic decay constants etc.,
which are yet poorly known (see for example \cite{R3}).

Currently there is impressive effort in this direction, and many new
facilities are under construction for studying the rare $B$--meson decays,
namely, symmetric and asymmetric $B$--meson factories at Cornell, KEK and
SLAC. Progress is also being made in hadronic environment at HERA--B and
there are some plans for $TeV$--$B$ and $LHC$--$B$. These machines will serve to
measure the processes, for which SM predicts very small Branching Ratios.
Among the rare decays, the flavor changing decays of the $B$--meson which
proceed via electroweak penguins, are of special interest due to their
relative cleanliness and their sensitivity to the new physics. The rare 
$B \rar \tau^+ \tau^- \gamma$ decay belongs to this category.

From helicity arguments it is clear that the matrix element of $B \rar
\ell^+ \ell^- ~(\ell = e,~\mu)$ decay will be proportional to the lepton mass
and therefore the corresponding Branching Ratios will be strongly
suppressed. Note that in SM, ${\cal B}(B \rar e^+ e^-) \simeq 4.2 \times
10^{-14}$ and ${\cal B}(B \rar \mu^+ \mu^-) \simeq 1.8 \times 10^{-9}$.
It is well known however that the the ${\cal B}(B \rar \tau^+ \tau^-) \simeq 
8 \times 10^{-7}$ in SM \cite{R4}, and thus this decay can be measured in future
$B$--factories with high enough efficiency.

When photon is emitted in addition to the lepton pair, no helicity suppression exists, 
and a "large" Branching Ratio is expected. Indeed in \cite{R5} it was shown that
the ${\cal B}(B \rar e^+ e^- \gamma) \simeq 2.35  \times 10^{-9}$. For  
$B \rar \ell^+ \ell^- \gamma  ~(\ell = e,~\mu)$, the contributions of the
diagrams, where photon is radiated from any charged internal line, can safely be
neglected, as they are strongly suppressed by a factor $m_b^2/m_W^2$ in the
Wilson coefficients (see \cite{R5}). Moreover, it follows from helicity arguments that, 
the contributions of the diagrams where a photon is emitted from the final charged lepton 
lines, must be proportional to the lepton mass $m_\ell~(\ell =e,~\mu)$, and
hence they are negligible as well.
Therefore in $B \rar \ell^+ \ell^- \gamma ~ (\ell = e,~ \mu)$, the main 
contribution should come from the diagrams, where photon is emitted from the
initial quarks.

In $B \rar \tau^+ \tau^- \gamma$ decay, the situation becomes very
different. In this case, we cannot neglect the contribution of the
diagrams, where photon is radiated from the final $\tau$--leptons, since the
mass of the $\tau$--lepton is not so much smaller than that of the
$B$--meson. So, in $B \rar \tau^+ \tau^- \gamma$ decay comparable
contributions
come from diagrams where photon is radiated both from initial and final
fermions. These contributions can give essential information about the
relative roles of the strong and electroweak interactions.

In this work we investigate the $B \rar \tau^+ \tau^- \gamma$ decay, and the
paper is organized as follows. In Section 2 we give the necessary theoretical
framework for the $B \rar \tau^+ \tau^- \gamma$ decay. Section 3 is devoted to
the numerical analysis and the discussion of the results.
In Appendix the detailed description of the cancellation of the infrared
(IR) singularities, is given.
 
\section{Theoretical framework for the $B \rar \tau^+ \tau^- \gamma$ decay.}

The matrix element for the $b \rar s \tau^+ \tau^- \gamma$ decay can be
obtained from that of the $b \rar s \tau^+ \tau^-$. It is well known that the
short distance contributions to $b \rar s \tau^+ \tau^-$ decay comes from the
box, Z-- and photon--mediated penguins. Thus, in the SM, $QCD$--corrected
amplitude for $b \rar s \tau^+ \tau^-$ can be written as \cite{R6, R7}.

\bea
{\cal M} &=& \frac{\alpha G_F}{2 \sqrt{2}\, \pi} V_{tb} V_{ts}^*
\Bigg{\{} C_9^{eff} \bar s \gamma_\mu (1-\gamma_5) b \, \bar \tau \gamma_\mu \tau +
C_{10}\bar s \gamma_\mu (1-\gamma_5) b \, \bar \tau \gamma_\mu \gamma_5
\tau~ \nnb \\
&& -~ 2 C_7 \frac{m_b}{p^2} \bar s i \sigma_{\mu \nu} p_\nu (1+\gamma_5) b \,
\bar \tau \gamma_\mu \tau ~
\Bigg{\}}~.
\eea

In Eq. (1) $p$ is the momentum transfer, and the mass of the strange quark is
neglected, $V_{ij}$'s are the corresponding elements of the CKM matrix. The
analytical expression of all Wilson coefficients $C_9^{eff}$, $C_{10}$ and
$C_7$ can be found in \cite{R6, R7}. In order to obtain the matrix element
for $b \rar s \tau^+ \tau^- \gamma$, it is necessary to attach photon to any
charged internal, as well as external line. Contributions of
the diagrams with photon attached to the any charged internal line,
are strongly suppressed and therefore we shall neglect these in the
following discussions. Thus, as explained previously the main contributions
to the $b \rar s \tau^+ \tau^- \gamma$ decay comes from diagrams, when
photon is radiated from initial and final fermions. 

When a photon is attached to the initial quark lines, the corresponding
matrix element for the $B \rar \tau^+ \tau^- \gamma$ decay can be written as

\bea
{\cal M}_1 &=& \la \gamma \ve {\cal M}\ve B \ra = \nnb \\                           
&& \frac{\alpha G_F}{2 \sqrt{2} \, \pi} V_{tb} V_{ts}^* \Bigg{\{} C_9^{eff} 
\bar \tau \gamma_\mu \tau \la \gamma \ve \bar s \gamma_\mu (1-\gamma_5) b \ve 
B \ra~ \nnb \\                                                        
&& +~ C_{10} \bar \tau \gamma_\mu \gamma_5 \tau \la \gamma \ve \bar s
\gamma_\mu(1-\gamma_5) b \ve B\ra~\nnb \\                 
&& -~ 2 C_7 \frac{m_b}{p^2} \la \gamma \ve \bar s i \sigma_{\mu \nu} p_{\nu} (1+
\gamma_5) b \ve B \ra \bar \tau \gamma_\mu \tau\Bigg{\}}~.                             
\eea
 
These matrix elements can be written in terms of the two independent, gauge
invariant, parity conserving and parity violating form factors \cite{R5,R8}:

\bea
\la \gamma \ve \bar s \gamma_\mu ( 1- \gamma_5) b \ve B \ra &=&
\frac{e}{m_B^2} \Bigg{\{}                                                      
\epsilon_{\mu \alpha \beta \sigma} \epsilon_\alpha^* p_\beta q_\sigma   
\, g(p^2)~ \nnb \\                                    
&& + ~i \left[ \epsilon_\mu^* (p q ) - (\epsilon^* p) q_\mu\right] \, f(p^2)
\Bigg{\}}~, \nnb \\                                              
\la \gamma \ve \bar s i \sigma_{\mu \nu} p_\nu (1+ \gamma_5) b \ve
B \ra &=& \frac{e}{m_B^2} \Bigg{\{} \epsilon_{\mu \alpha \beta \sigma}
\epsilon_\alpha^*
p_\beta q_\sigma \, g_1(p^2) ~\nnb \\                 
&& +~ i \left[ \epsilon_\mu^* (p q) - (\epsilon^* p ) q_\mu \right] \, f_1(p^2)
\Bigg{\}}~.
\eea
Here $\epsilon_\mu$ and $q_\mu$ are the four vector polarization and four
momentum of the photon, respectively, and $p$ is the momentum transfer.
Substituting Eq. (3) in (4), for the matrix element ${\cal M}_1$ (structure
dependent part) we get
\bea
{\cal M}_1 &=& \frac{\alpha G_F}{2 \sqrt{2} \, \pi} V_{tb} V_{ts}^* e 
\Bigg{\{} \epsilon_{\mu \alpha \beta \sigma} \epsilon_\alpha^* p_\beta
q_\sigma \left[ A \, \bar \tau \gamma_\mu \tau + C \, \bar \tau \gamma_\mu 
\gamma_5 \tau \right] ~ \nnb \\
&& +~  i \left[ \epsilon_\mu^* (p q) - (\epsilon^* p ) q_\mu \right] 
\left[ B \, \bar \tau \gamma_\mu \tau + D  \, \bar \tau \gamma_\mu \gamma_5
\tau \right] \Bigg{\}}~,
\eea
where 
\bea
A &=& \frac{1}{m_B^2} \left[ C_9^{eff} \,g(p^2) - 
2 C_7 \frac{m_b}{p^2} \, g_1(p^2) \right]~, \nnb \\
B &=& \frac{1}{m_B^2} \left[ C_9^{eff} \,f(p^2) - 
2 C_7 \frac{m_b}{p^2} \, f_1(p^2) \right]~,\nnb \\
C &=& \frac{C_{10}}{m_B^2} \,g(p^2)~,\nnb \\
D &=& \frac{C_{10}}{m_B^2} \,f(p^2)~.
\eea

When a photon is radiated from the final $\tau$--leptons, the corresponding
matrix element is (Bremstrahlung part)

\bea
{\cal M}_2 =  \frac{\alpha G_F}{2 \sqrt{2} \, \pi} V_{tb} V_{ts}^* e
i f_B C_{10} 2 m_\tau \left[ \bar \tau \left(
\frac{\not\!\epsilon \! \not\!P_B}{2 p_1 q} - 
\frac{\not\!P_B \! \not\!\epsilon}{2 p_2  q} \right) \gamma_5 \tau \right]~,
\eea
where $P_B$ is the momentum of the $B$--meson. 
In obtaining this expression we have used
\bea
\la 0 \ve \bar s \gamma_\mu \gamma_5 b \ve B \ra &=& -~i f_B P_{B\mu}~, \nnb \\
\la 0 \ve \bar s \sigma_{\mu\nu} (1+\gamma_5) b \ve B \ra &=& 0~,
\eea
and the conservation of the vector current. 

Finally, the total matrix element for
the $B \rar \tau^+ \tau^- \gamma$ decay is obtained as a sum of the 
${\cal M}_1$ and ${\cal M}_2$:
\bea
{\cal M} = {\cal M}_1 + {\cal M}_2~.
\eea
The square of the matrix element, summed over the spins of the
$\tau$--leptons and the polarization of the photon, can be written as

\bea
\ve {\cal M} \ve^2 = \ve {\cal M}_1 \ve^2 + \ve {\cal M}_2 \ve^2 + 
2 \,{\rm Re} \left( {\cal M}_1 {\cal M}_2^*\right)~,
\eea
where
\newpage
\bea
\vel {\cal M}_1 \ver^2 &=& \vel \frac{\alpha G_F}{2 \sqrt{2} \, \pi} 
V_{tb} V_{ts}^* \ver^2 4 \pi \alpha \, 
\Bigg{\{} 8 \, {\rm Re} \ga B^* C + A^* D \dr p^2 \ga p_1 q - p_2 q \dr
\ga p_1 q + p_2 q \dr ~ \nnb \\
&& +~ 4 \left[ \vel C \ver^2 + \vel D \ver^2 \right] 
\left[ \ga p^2 - 2 m_\tau^2 \dr \ga \ga p_1 q \dr^2 + \ga p_2 q \dr^2
\dr - 4 m_\tau^2 \ga p_1 q \dr \ga p_2 q \dr \right]~  \nnb \\
&& +~ 4 \left[ \vel A \ver^2 + \vel B \ver^2 \right]
\Big[ \ga p^2 + 2 m_\tau^2 \dr \ga \ga p_1 q \dr^2 + \ga p_2 q \dr^2
\dr~ \nnb \\
&& + ~ 4 m_\tau^2 \ga p_1 q \dr \ga p_2 q \dr \Big] \Bigg{\}}~, 
\\ \nnb\\ \nnb\\
2\, {\rm Re} \ga {\cal M}_1 {\cal M}_2^* \dr &=&
-~\vel \frac{\alpha G_F}{2 \sqrt{2} \, \pi} 
V_{tb} V_{ts}^* \ver^2 4 \pi \alpha \,
\Bigg{\{} 16 \, C_{10} f_B m_\tau^2 \Bigg[ {\rm Re}(A) \, 
\frac{ \ga p_1 q +  p_2 q \dr^3}{\ga p_1 q \dr \ga p_2 q \dr}~ \nnb \\
&& +~ {\rm Re}(D) \,
\frac{ \ga p_1 q +  p_2 q \dr^2  \ga p_2 q -  p_1 q \dr}
{\ga p_1 q \dr \ga p_2 q \dr} \Bigg] \Bigg{\}}~, \\ \nnb \\ \nnb \\
\vel {\cal M}_2 \ver^2 &=& -~\vel \frac{\alpha G_F}{2 \sqrt{2} \, \pi}
V_{tb} V_{ts}^* \ver^2 4 \pi \alpha \,
\Bigg{\{} - 32 + 8 \, \frac{m_\tau^2}{\ga p_1 q \dr^2} \ga p^2 + 2 p_2 q \dr
~\nnb \\
&& +~ \frac{16}{p_1 q} \left[ 3 m_\tau^2 - p^2 - p_2 q \right] + 
8  \, \frac{m_\tau^2}{\ga p_2 q \dr^2}  \ga p^2 + 2 p_1 q \dr ~\nnb \\
&& +~ \frac{16}{p_2 q} \left[ 3 m_\tau^2 - p^2 - p_1 q \right] + 
8  \, \frac{p^2}{\ga p_1 q \dr \ga p_2 q \dr} \left[ 2  m_\tau^2 - p^2 \right]
\Bigg{\}}~.
\eea
Here $p_1,~p_2$ are momenta of the final $\tau$--leptons, and $q$ is the photon
momentum. The quantity $\vel {\cal M} \ver^2$ depends only on the scalar
products of the momenta of the external particles. In the rest frame of the
$B$--meson, all these scalar products are fixed, if the photon energy
$E_\gamma$ and the lepton energy $E_1$ are specified. The Dalitz boundary is
given as
\bea
0 \leq E_\gamma \leq \frac{m_B^2 - 4 m_\tau^2}{2 m_B}~,
\eea
\bea
\frac{m_B - E_\gamma}{2} - \frac{E_\gamma}{2} 
\sqrt{1 - \frac{4 m_\tau^2}{m_B^2 - 2 m_B E_\gamma}} \leq E_1 \leq
\frac{m_B - E_\gamma}{2} + \frac{E_\gamma}{2} 
\sqrt{1 - \frac{4 m_\tau^2}{m_B^2 - 2 m_B E_\gamma}}~.
\eea
The $\vel {\cal M}_1 \ver^2$ term is completely infrared--free; the
interference term has an integrable infrared singularity. Thus infrared
divergence appears only in $\vel {\cal M}_2 \ver^2$. The infrared
singularity originates in the Bremstrahlung processes, when photon is soft. 
It is clear that
in this limit, the $B \rar \tau^+ \tau^- \gamma$ decay cannot be
distinguished from $B \rar \tau^+ \tau^-$. Therefore both processes must be
considered together in order to obtain finite result for the decay rate. In
the Appendix we show that IR singular terms in $\vel {\cal M}_2 \ver^2$
exactly cancel the $O(\alpha)$ virtual correction in $B \rar \tau^+
\tau^-$ amplitude. 

In this work, our point of view is slightly different from the standard
description. Namely, we consider the Bremstrahlung process as a different
process but not as the $\alpha$ correction to the $B \rar \tau^+
\tau^-$ decay. In other words, we consider the photon in $B \rar \tau^+
\tau^- \gamma$ as a hard photon. Therefore, in order to obtain the decay
width of the $B \rar \tau^+ \tau^-$ + (hard photon), we must impose a cut
on the photon energy, which will correspond to the experimental cut imposed
on the minimum energy for detectable photon. We require the energy of the
photon to be larger than $50~MeV$, i.e., $E_\gamma \geq a\, m_B$, where
$a \geq 0.01$.

After integrating over the phase space, and taking into account the cut for
the photon energy, for the decay rate we get,

\bea
\Gamma &=& \vel \frac{\alpha G_F}{2 \sqrt{2} \, \pi} V_{tb} V_{ts}^* \ver^2 \,
\frac{\alpha}{\ga 2 \, \pi \dr^3}\, m_B^5 \pi \nnb \\
&\times& \Bigg{\{} \frac{1}{12} \, \int_\delta^{1-4 r} x^3 \,dx \,
\sqrt{1-\frac{4 r}{1 - x}} \, m_B^2 \Big[ \ga \vel A \ver^2 + \vel B \ver^2
\dr \ga 1- x+ 2 r \dr \nnb \\
&+& \ga \vel C \ver^2 + \vel D \ver^2 \dr \ga 1- x - 4 r \dr \Big] \nnb \\
&-& 2 C_{10} f_B r \int_\delta^{1-4 r} x^2 \,dx \, {\rm Re} \ga A \dr \,
{\rm ln} 
\displaystyle{\frac{1 + \sqrt{1-\displaystyle{\frac{4 r}{1 - x}}}}
{1 -  \sqrt{1-\displaystyle{\frac{4 r}{1 - x}}}}}
\nnb \\
&-& 4 \vel f_B C_{10} \ver^2 r \, \frac{1}{m_B^2} \, \int_\delta^{1-4 r} dx
\Bigg[ \ga 2 + \frac{4 r }{x} -\frac{2}{x} -x \dr\,
{\rm ln}
\displaystyle{\frac{1 + \sqrt{1-\displaystyle{\frac{4 r}{1 - x}}}}
{1 -  \sqrt{1-\displaystyle{\frac{4 r}{1 - x}}}}}
\nnb \\
&+& \frac{2}{x} \ga 1-x \dr \, \sqrt{1-\frac{4 r}{1 - x}}\, \Bigg] \Bigg{\}}~, 
\eea
where $x=\displaystyle{\frac{2 E_\gamma}{m_B}}$ is the dimensionless photon energy,
$r=\displaystyle{\frac{m_\tau^2}{m_B^2}}$ and $\delta=2 a$, satisfying
\bea
\delta \leq x \leq 1 - \frac{4 m_\tau^2}{m_B^2}~. \nnb
\eea
From Eq. (15) it follows that for calculating the decay width we need
explicit forms of the form factors $g,~f,~g_1$ and $f_1$. These form factors
are calculated in the framework of light--cone $QCD$ sum rules in \cite{R4}
(see also \cite{R8}), and their $p^2$ dependences, to a very good accuracy,
can be represented in the following dipole forms,
\bea
g_(p^2) &=& \frac{1~GeV}{(1-\displaystyle{\frac{p^2}{5.6^2})^2}}~,
~~~~~~~~~~
f_(p^2) = \frac{0.8~GeV}{(1-\displaystyle{\frac{p^2}{6.5^2})^2}}~,
 \nnb \\ \nnb  \\ \nnb \\ 
g_1(p^2) &=& \frac{3.74~GeV^2}{(1-\displaystyle{\frac{p^2}{40.5})^2}}~, 
~~~~~~~~~
f_1(p^2) = \frac{0.68~GeV^2}{(1-\displaystyle{\frac{p^2}{30})^2}}~,
\eea
which we will use in the numerical analysis.

\section{Numerical analysis and discussion}

For the input parameters, which enter into the expression for the decay
width we use the following values: $m_b=4.8~ GeV,~m_c=1.35~
GeV,~m_\tau=1.78~ GeV,~m_B=5.28~ GeV,~\vel V_{tb} V_{ts^*}
\ver=0.045$. We use the pole form of the form factors given in Eq. (16). For
$B$--meson life time we take $\tau\ga B_s \dr = 1.64 \times 10^{-12}~s$
\cite{R9}. The value of the Wilson coefficients $C_7 \ga m_b \dr$ and
$C_{10}\ga m_b \dr$, to the leading logarithmic approximation, are (see for
example \cite{R6, R7}):
\bea
C_7 \ga m_b \dr = -\,0.315,~~~~~C_{10}\ga m_b \dr = -\,4.6242~.
\eea

The expression $C_9^{eff}\ga m_b \dr$ for the $b \rar s$ transition, in the
next--to--leading--order approximation, is given as
\bea
C_9^{eff}(m_b) &=& C_9(m_b) + 0.124 w(\hat{s}) + g(\hat{m}_c,\hat{s})( 3 C_1 + C_2 + 3
C_3 + C_4 + 3 C_5 + C_6)  \nnb \\
&& - \frac{1}{2} g(\hat{m}_q,\hat{s})(C_3 + 3 C_4)
-\frac{1}{2} g(\hat{m}_b,\hat{s})(4 C_3 + 4 C_4 + 3 C_5 + C_6)  \nnb \\
&& + \frac{2}{9} (3 C_3 + C_4 + 3 C_5 + C_6)~,
\eea
with 
\bea
&& C_1 = -\,0.249, ~~~~~ C_2 = 1.108, ~~~~~ C_3 = 1.112 \times 10^{-2}, 
C_4 = -\,2.569 \times 10^{-2}, \nnb \\
&& C_5 = 7.4 \times 10^{-3}, ~~~~~ C_6 = -\, 3.144 \times 10^{-2},
~~~~~ C_9 = 4.227~,
\eea
where $\hat{m}_q = m_q/m_b,~\hat{s} = p^2/m_b^2$. The value of $C_9^{eff}$
for the $b \rar d$ transition can be obtained by adding to Eq. (18) the term
\bea
\lambda_u \left[ g \ga \hat{m}_c,\hat{s} \dr - g \ga \hat{m}_d,\hat{s}\dr
\right] \ga 3 C_1 + C_2 \dr~, \nnb 
\eea
where
\bea
\lambda_u = \frac{V_{ub} V_{ud}^*}{V_{tb} V_{td}^*} ~, \nnb
\eea
and replacing $V_{tb} V_{ts}^*$ in Eq. (15) by $V_{tb} V_{td}^*$.
In Eq. (18), $w(\hat{s})$ represents the one gluon correction to the matrix
element of operator $O_9$, and its explicit form can be found in \cite{R6, R7}, 
while the function $g(\hat{m}_q,\hat{s})$ arises from the 
one--loop contributions of the four quark operators $O_1$--$O_6$, i.e., 
\bea
\lefteqn{
g(\hat{m}_q,\hat{s}')= - \frac{8}{9} {\rm ln}(\hat{m}_q) + \frac{8}{27} +
\frac{4}{9} y_q - \frac{2}{9} (2 + y_q) \sqrt{\vel 1-y_q \ver} } \nnb \\ \nnb \\ 
&& \times \Bigg{\{} \Theta(1-y_q) 
\ga {\rm ln} \frac{1+\sqrt{1-y_q}}{1-\sqrt{1-y_q}} - i
\pi \dr + \Theta(y_q-1) {\rm arctg} \frac{1}{\sqrt{y_q-1}} \Bigg{\}}~, \nnb 
\eea
where $y_q = 4 \hat{m}_q^2 / \hat{s}'$, and $\hat{s}'= p^2/m_b^2$.

For a more complete analysis of the $B \rar \tau^+ \tau^- \gamma$ decay, one
has to take into account the long distance contributions. For this aim it is
necessary to make the following replacement
\bea
g(\hat{m}_c,\hat{s}') \rar g(\hat{m}_c,\hat{s}') - \frac{3 \pi}{\alpha^2}
\sum_{V=J/\psi , \psi'} \frac{\hat{m}_V {\rm B}( V \rar \tau^+ \tau^-)
\hat{\Gamma}_{tot}^V}{\hat{s}' - \hat{m}_V^2 + i \hat{m}_V
\hat{\Gamma}_{tot}^V}~,
\eea
where $\hat{m}_V = m_V/m_b$, $\hat{\Gamma}_{tot} = \Gamma /m_b$.

Our results for the Branching Ratio ${\cal B}(B \rar  \tau^+ \tau^- \gamma)$ for two
different values of the cut ($\delta = 0.01$ and $\delta = 0.02$) are presented in
Table 1.  
  

\begin{table}[h]
\begin{center}
\begin{tabular}{|c|c|c|c|c|}
\hline
\multicolumn{1}{|c|}{\aaa }
&\multicolumn{2}{|c|}{{\bf Short Distance}}
&\multicolumn{2}{|c|}{{\bf Short and Long Distance}} \\
\multicolumn{1}{|c|}{ }&\multicolumn{2}{|c|}{{\bf
Contributions to the}}&\multicolumn{2}{|c|}{{\bf Contributions to the}} \\
\multicolumn{1}{|c|}{ }&\multicolumn{2}{|c|}{\dol{\bf  
Branching Ratio}}&\multicolumn{2}{|c|}{{\bf Branching Ratio}} \\
\hline\hline        
             &\bos $ \delta = 0.01 $ & $ \delta = 0.02 $ 
& $ \delta = 0.01  $ & $  \delta = 0.02 $ 
\\  \hline
{\bf Structure}     & \aaa $4.19 \times 10^{-9}$ & $4.19 \times 10^{-9}$ 
& $9.95 \times 10^{-9}$ & $7.68  \times 10^{-9}$    \\
\dol{\bf  dependent part} &          &          &          & \\ 
\hline      
{\bf Bremstrahlung} & \aaa $4.11 \times 10^{-9}$ & $3.16 \times 10^{-9}$ 
& $4.11 \times 10^{-9}$ & $3.16 \times 10^{-9}$ \\
\dol {\bf part} &          &          &          & \\
\hline      
{\bf Interference}  & \aaa $1.24 \times 10^{-9}$ & $1.23 \times 10^{-9}$ 
& $1.16 \times 10^{-9}$ & $1.16 \times 10^{-9}$ \\
\dol {\bf part} &          &          &          &  \\
\hline
{\bf Total}         & \bos $9.54 \times 10^{-9}$ & $8.59 \times 10^{-9}$ 
& $1.52 \times 10^{-8}$ & $1.20 \times 10^{-8}$ \\
\hline      
\end{tabular}
\vskip 0.3 cm
\caption{}
\end{center}
\end{table} 
            

Note that, when only short distance effects are taken into
consideration, the structure--dependent and the Bremstrahlung parts give,
more or less, comparable contributions. However, if long distance contributions are 
also considered in addition to the short ones, the
structure--dependent part contributes more as compared to that of the Bremstrahlung
part. This is due to the fact that, the structure dependent part contains 
$(J/\psi , \psi')$ resonance contributions in the latter case (see Eq. (20)).

From these results it follows that there is a good chance for detecting $\tau$ 
lepton decay in the future $B$--meson factories, provided that the efficiency is 
$\sim$ 1/3.

\newpage
\section*{Figure Captions}
{\bf 1.}  Dependence of the Differential Branching Ratio for the 
$B \rar \tau^+ \tau^- \gamma$ decay on the dimensionless variable 
$x=\displaystyle{\frac{2 E_\gamma}{m_B}}$, for the value of the cut $\delta =
0.01$ imposed on the photon energy. In this figure, 
the curve with the sharp peak represents the long distance contributions. \\ \\
{\bf 2.} Same as Fig. 1, but for the cut value $\delta =0.02$.

\begin{figure}
\vspace{25.cm}
    \includegraphics{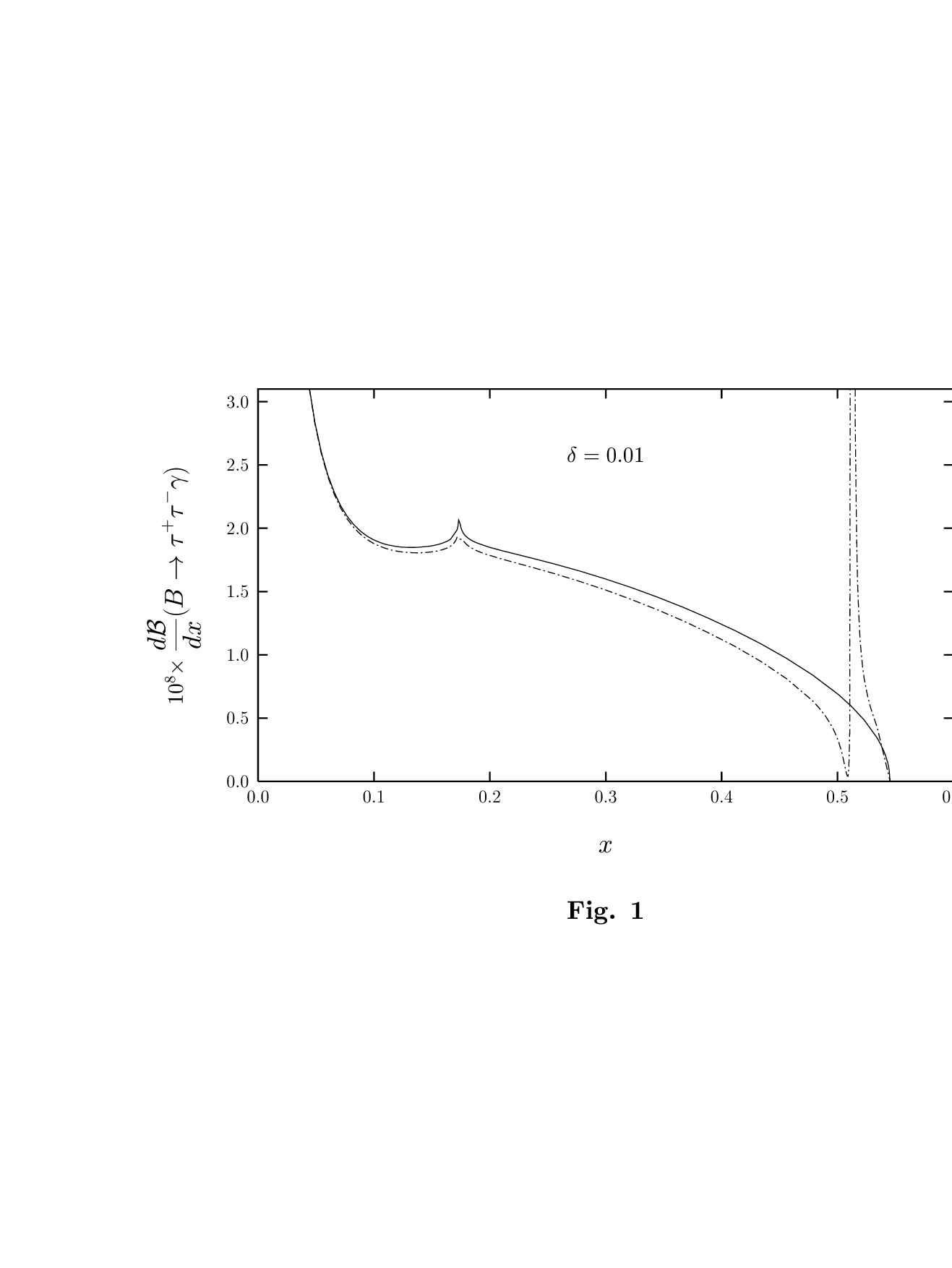}
    \includegraphics{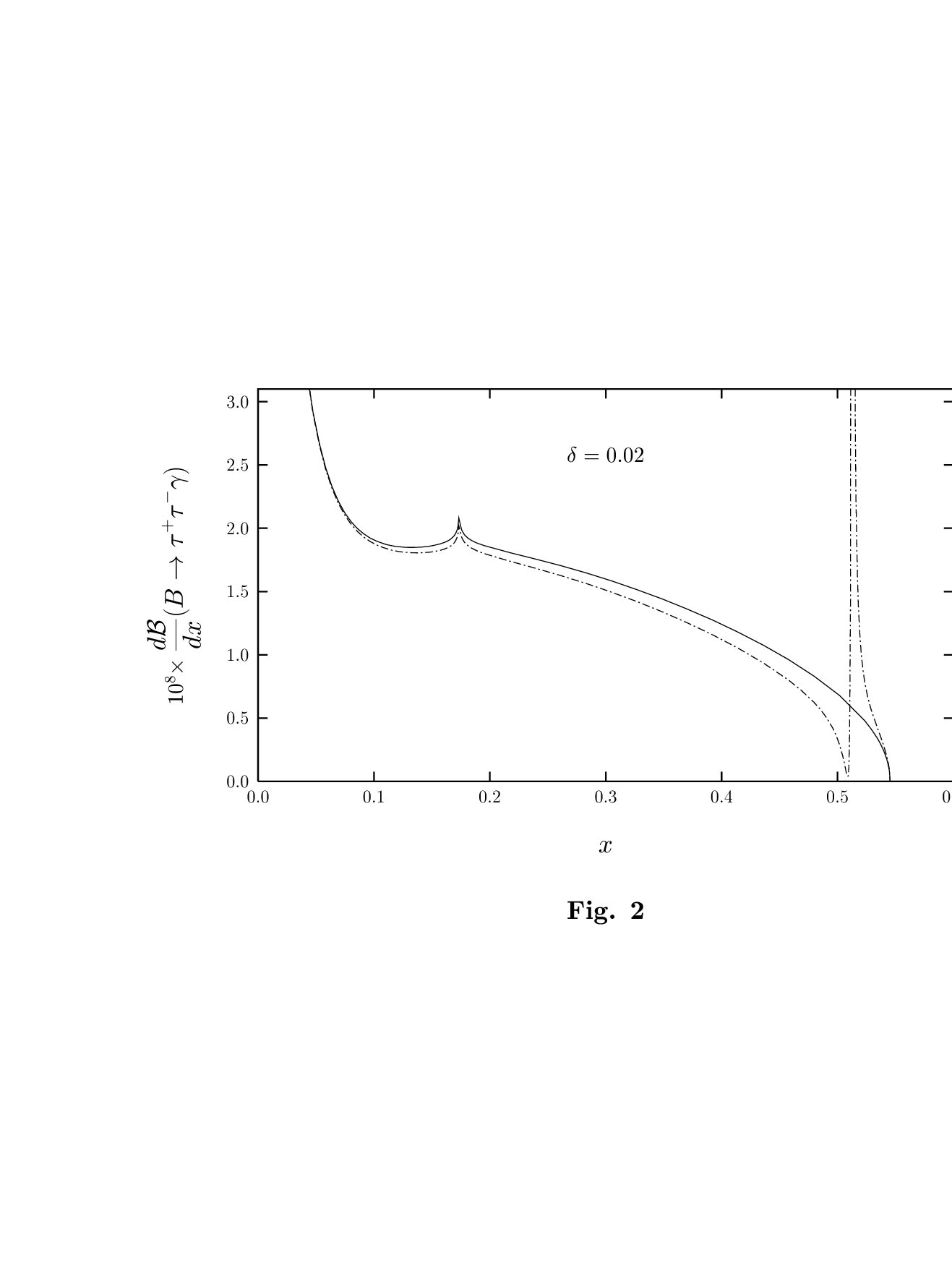}
    \vspace{-4.0cm}
\end{figure}

\newpage
\bAPP{A}{Cancellation of the infrared divergences}
In this section we will show the explicit cancellation between the infrared
singularities arising from the Bremstrahlung of a soft photon in the $B \rar
\tau^+ \tau^- \gamma$ rate and in the $O(\alpha)$ virtual corrections to
the  $B \rar \tau^+ \tau^-$ amplitude. 

As we have explained in Section 2, we are not going to include $O(\alpha)$
virtual corrections to the $B \rar \tau^+ \tau^-$ amplitude in the
calculation of the rate of $B \rar \tau^+ \tau^- \gamma$. The Bremstrahlung
process here is not considered as an $O(\alpha)$ correction to the $B \rar
\tau^+ \tau^-$ amplitude, but as a different process, namely the decay of a
$B$--meson into $\tau$--lepton pair plus a hard photon. In order to
calculate the physical rate of interest, we just have to impose a cut on the
energy of photon, which will correspond to the experimental cut imposed on
the minimum energy for detectable photon.

In this sense our approach is very similar to the one given in \cite{R11},
in studying $B_s \rar X_s \gamma \gamma$ decay. Therefore, in the present
Appendix we will consider only those aspects of the discussion which are
necessary to show the cancellation of the infrared singularities. 
For removing the infrared singularities we will use the method presented in
\cite{R12}, where the main idea is to use the dimensional regularization
method, i.e., to replace

\baeq
\frac{d^3 q}{2\,E_\gamma \,\ga 2 \pi \dr^3} \rar 
\frac{d^{(n-1)} q}{2 \vel q \ver \ga 2 \pi \dr^{(n-1)}}~,
\eaeq
where $q=\ga q^0,~q_i \dr$ is an $n$ dimensional light--like vector:
\bea
q^0 = \vel q \ver = \ga \vel q_1 \ver^2 +  \cdots  
+ \vel q_{n-1} \ver^2 \dr^{1/2}~. \nnb
\eea
All calculations are performed in the rest frame of the $B$--meson. We
are interested in the situation when charged lepton is observed in the
energy interval $E_m - \Delta E \leq E \leq E_m$, where $E_m$ is that of
the charged lepton energy in the two--body decay $B \rar \tau^+ \tau^-$ and 
$\Delta E <\!\!< E_m$. We will retain terms of logarithmic and zeroth order
in $\Delta E$. It is clear that in this limit, only soft photons give
contributions.

Let us first consider the Bremstrahlung part. From $\vel {\cal M}_2 \ver^2$
it follows that only terms proportional to $1/ \ga p_1 q \dr^2,~1/ \ga p_2 q
\dr^2$ and $1/ \left[ \ga p_1 q \dr \ga p_2 q \dr \right]$ give 
infrared singularities (we
will omit the terms which give finite contributions to the to the decay
rate), i.e.,
\baeq\eqalign{
\vel {\cal M}_2 \ver^2_{IR} &= -~\vel \frac{\alpha G_F V_{tb} V_{ts}^* f_B
C_{10}}{2 \sqrt{2}\, \pi}\ver^2 \, m_\tau^2 4 \pi \alpha \, \cr
&\times \Bigg{\{} 8 m_\tau^2 \frac{p^2}{\ga p_1 q \dr^2} + 8 m_\tau^2 \frac{p^2}{\ga
p_2 q \dr^2} + 16 m_\tau^2 \left[ \frac{ \ga p_1 q \dr }{\ga p_2 q \dr^2} +
\frac{ \ga p_2 q \dr }{\ga p_1 q \dr^2}\right] \cr
&+ 16 m_\tau^2 \frac{p^2}{ \ga p_1 q \dr \ga p_2 q \dr } - 
8 \frac{p^4}{ \ga p_1 q \dr \ga p_2 q \dr } \Bigg{\}}~.
}\eaeq

Let us consider the first term in the bracket

\baeq\eqalign{
I &= \int \frac{d^3 \vec{p_1}}{2 E_1} \, \frac{d^3 \vec{p_2}}{2 E_2}\, 
\frac{d^{(n-1)} q}{2 \vel q \ver \ga 2 \pi \dr^{n-1}} \,
\frac{p^2}{\ga p_1 q \dr^2} \, \frac{1}{\ga 2 \pi \dr^2} 
\, \delta \! \ga p_B - p_1 - p_2 - q \dr \cr
&= \int \frac{d^3 \vec{p_1}}{2 E_1} \, \frac{d^{(n-1)} q}{2 \vel q \ver
\ga
2 \pi \dr^{n-1}} \, \frac{p^2}{\ga p_1 q \dr^2} \, \frac{1}{\ga 2 \pi \dr^2}
\, \delta \! \ga m_B^2 - 2 m_B E_1 - 2 m_B q^0 + 
2 E_1 q \ga 1 - \beta z \dr \dr  ~,
}\eaeq
where $\beta = \vel \vec{p_1} \ver /E_1$, and we choose the first axis along
the direction of $\vec{p_1}$. The integrand depends only on the angle
$\theta_1 = \theta$, so that we can immediately perform integration over the
other angles. Integration over all of the angular variables leads to the
following result
\baeq\eqalign{
\int d^{(n-1)} q \rar \frac{2 \pi^{n/2 - 1}}{\Gamma \ga n/2 -1 \dr }
\int d \vel q \ver q^{(n-2)} \int_{-1}^1 d z\, \ga 1 - z^2 \dr^{n/2-2}~,
}\eaeq
where $z = cos\theta$. 
Using (A.4), and performing integration over radial $d \vel q \ver$ we obtain
\baeq\eqalign{
I &= \frac{1}{2 \ga 2 \pi \dr^{n-1}} \, \frac{1}{\ga 2 \pi \dr^2} \,
\frac{2 \pi^{n/2-1}}{\Gamma \ga n/2-1 \dr} 
\int_{E_m - \Delta E}^{E_m} \frac{4 \pi \beta \, d E_1}{4 E_1} 
\int d z \,  \frac{\ga 1 - z^2 \dr^{n/2-1}}{\ga 1 - \beta z \dr^2} \cr
&\times \Bigg{\{} m_B^2 
\frac{ \ga m_B^2 - 2 m_B E_1 \dr^{n-5}}{ \left[2 m_B - 2 E_1 \ga 1 - \beta z
\dr \right]^{n-4}} -
2 m_B \frac{\ga m_B^2 - 2 m_B E_1 \dr^{n-4}}{ \left[2 m_B - 2 E_1 \ga 1 -
\beta z   
\dr \right]^{n-3}} \Bigg{\}}~.
}\eaeq
Second term gives finite contribution and therefore we will omit it. In our
case $E_m = m_B/2$.

Introducing next a new variable $t=E_m - E_1$ and expanding all terms in
Taylor series about $E_1=E_m$, we get
\baeq\eqalign{
I &= \frac{1}{2^{n+2} \pi^{n/2+1}} \, \frac{1}{\Gamma \ga n/2 -1 \dr}
\int dz \, \frac{ \ga 1 - z^2 \dr^{n/2-2}}{\ga 1-\beta_m z\dr^2} \,
\frac{\beta_m}{E_m} \cr
&\times \int_0^{\Delta E} dt \, \frac{m_B^{n+3} t^{n-5}}
{2 E_m^{n-4} \ga 1 + \beta_m z \dr^{n-4} }~.
}\eaeq
After integrating over $t$ we have
\baeq\eqalign{
I &= \frac{1}{2^{n+2} \pi^{n/2+1}} \, \frac{1}{\Gamma \ga n/2 -1 \dr}
\, \frac{\beta_m}{2} \cr
&\times \Bigg{\{} 
\ga \frac{1}{n-4} + {\rm ln} \Delta E \dr \int_{-1}^{1} d z \,
\frac{ \ga 1 - z^2 \dr^{n/2-2}}{\ga 1-\beta_m z\dr^2} \, 
\frac{1}{\ga 1 + \beta_m z \dr^{n-4} } \Bigg{\}}~.
}\eaeq
The last step is to expand (A.7) in a Laurent series about $n=4$ and perform
trivial integrations over $d x$ (retaining only $1/(n-4)$ and ${\rm ln} 
\Delta E$ terms). Calculating in a similar manner all the other terms, we get (we
retain only infrared divergent terms and those that are proportional to
${\rm ln} \Delta E$)
\baeq\eqalign{
\Gamma_{\rm IR} &= -~ \vel  \frac{\alpha G_F V_{tb} V_{ts}^* f_B
C_{10}}{2 \sqrt{2}\, \pi}\ver^2 m_\tau^2 m_B^2 \frac{\alpha}{4} \beta_m
\pi^{-n/2} \frac{1}{2 m_B} \cr
&\times \Bigg{\{} \ga \frac{1}{n-4} + {\rm ln} \Delta E \dr 
\left[ 8 + 4 \ga 1 + \beta_m^2 \dr \right] \, \frac{1}{\beta_m} \,
{\rm ln} \frac{1-\beta_m}{1+\beta_m} + \cdots \Bigg{\}}~.
}\eaeq

Now let us calculate $O \ga \alpha \dr$ virtual corrections to the $B \rar
\tau^+ \tau^-$ decay. The matrix element for the $B \rar \tau^+ \tau^-$
decay with virtual corrections can be represented as 
\baeq\eqalign{
{\cal M} = {\cal M}_0 \left\{ 1 + 4 \pi \alpha {\rm K} + 
\ga Z_m -1 \dr \right\}~, \nnb 
}\eaeq
where
\bea                    
{\cal M}_0 = \frac{\alpha G_F}{2 \sqrt{2} \, \pi} 
f_B C_{10} V_{tb} V_{ts}^* 2 m_\tau \, \bar \tau \gamma_5 \tau ~, \nnb
\eea
is the matrix element for $B \rar \tau^+ \tau^-$ decay without virtual
corrections, K denotes $O \ga \alpha \dr$ corrections due to the photon
exchange (vertex) to the ${\cal M}_0$, and the last term corresponds to the
wave function renormalization. Note that all the calculations were performed
in the Landau gauge. The matrix element square with summation over spins of
the final particles
is given as
\baeq\eqalign{
\vel {\cal M} \ver^2 = \vel {\cal M}_0 \ver^2 \Big{\{}
1 + 4 \pi \alpha 2 {\rm Re(K)} + 2 \ga Z_m -1 \dr \Big{\}}~.
}\eaeq
In the Landau gauge, fermion wave function renormalization constant is given
as
\baeq\eqalign{
Z_m - 1 = \frac{\alpha}{4 \pi} \ga \frac{6}{n-4} - 4 \dr~.
}\eaeq
After standard calculation for the infrared singular part for the virtual
corrections, we finally get
\baeq\eqalign{
\Gamma_{\rm IR} &= \vel  \frac{\alpha G_F V_{tb} V_{ts}^* f_B
C_{10}}{2 \sqrt{2}\, \pi}\ver^2 m_\tau^2 m_B \frac{\alpha \beta}{4 \pi^2}
\cr
&\times \Bigg{\{} \frac{1}{n-4} \left[ 4 + \frac{2}{\beta} \ga 1 +
\beta \dr^2 \right] {\rm ln} \frac{1-\beta}{1+\beta} + \cdots \Bigg{\}}~.
}\eaeq
From Eqs. (A.8) and (A.12) we see that the infrared singular terms from
Bremstrahlung and vertex corrections exactly cancel each other, and the
decay rate is infrared--free. 
\eAPP

\end{document}